# Optical and electrical probing of plasmonic metal-molecule interactions

Andrei Stefancu*[1], Wenxuan Tang*[2], Ming Fu[2], Jordan Edwards[3], Naomi J. Halas[4], Ross C. Schofield[2], Toby Severs Millard[2], Peter Nordlander[4], Johannes Lischner[3], Pilar Carro[5], Rupert Oulton*[2], Emiliano Cortes*[1]

1 Nanoinstitute Munich, Faculty of Physics, Ludwig-Maximilians-Universität (LMU), Munich, Germany

2 Blackett Laboratory, Imperial College London, London, UK

3 Department of Materials, Imperial College London, South Kensington Campus, LondonSW7 2AZ, United Kingdom

4 Department of Electrical and Computer Engineering, Department of Chemistry, Department of Physics and Astronomy, Department of Material Science and Nanoengineering, and Laboratory for Nanophotonics, Rice University, Houston, Texas 77005, United States, Technical University of Munich (TUM) Institute for Advanced Study (IAS), Garching, Germany

5 Department of Chemistry, Faculty of Sciences, University of Laguna, Institute of Materials and Nanotechnology, 38200 La Laguna, Spain

**Abstract.** Plasmonic nanostructures enable efficient light-to-energy conversion by concentrating optical energy into nanoscale volumes. A key mechanism in this process is chemical interface damping (CID), where surface plasmons are damped by adsorbed molecules, enabling the transfer of charge to adsorbed molecules. In this study, we investigate the relationship between CID and adsorbate-induced changes in DC electrical resistivity for four molecular adsorbates—adenine, 4-aminothiophenol (ATP), biphenyl thiol (BPT), and 1-dodecanethiol (DDT)—on gold surfaces. Our results reveal two distinct CID regimes. BPT causes CID via direct electronic transitions to the lowest unoccupied molecular orbital (LUMO), which is centered at $\sim 2\ eV$ above the Fermi level and can be resonantly excited by the plasmon. This mechanism is dependent on plasmon energy. In contrast, ATP, adenine and DDT lead to plasmon damping through inelastic electron scattering at the metal-molecule interface. This regime shows a weaker dependency on plasmon energy since it does not involve resonant electron excitation between hybridized metal-molecule states. This same mechanism contributes to adsorbate-induced changes in DC resistivity, suggesting that resistivity measurements can serve as a probe of plasmonic energy transfer, as highlighted by the good correlation between the two effects. These findings provide new insights into the microscopic origins of plasmon damping and offer a unified framework for understanding metal–adsorbate energy transfer.

## Introduction

Plasmonic nanostructures have emerged as promising systems for efficient light harvesting and light-to-chemical-energy conversion, owing to their unique optical properties.[1-7] At the core of these properties are surface plasmon resonances (SPRs), coherent oscillations of the free metal electrons induced by resonant light. SPRs enable light absorption cross-sections that exceed the physical dimensions of the nanostructures by several orders of magnitude.[8] These resonances play a crucial role in driving chemical transformations in adsorbed molecules through charge and energy transfer.[9-10]

SPRs decay via two primary pathways: radiative decay, in which the plasmon emits a photon at the SPR frequency, and non-radiative decay, which generates electron-hole (e-h) pairs in the metal.[11-12] The energy partitioning between these pathways is determined by the dielectric constant of the material and its geometry and size. For plasmon-driven chemistry, non-radiative decay is particularly significant, as it facilitates plasmon-induced charge transfer to adsorbed molecules. Other mechanisms also exist for non-radiative plasmon-molecule energy transfer, such as plasmon-induced resonant energy transfer, the analogue of Forster resonance energy transfer.[13-14]

The transfer of energetic charge carriers (hot carriers) generated from SPRs to adsorbed molecules can take place through two mechanisms: direct charge transfer, also called chemical interface damping (CID), and a three-step hot electron mechanism. In the three-step process, SPRs decay by generating hot electrons in the metal from energy states near the Fermi level, $E_F$, to unoccupied states at $E_F + \hbar\omega_{SPR}$.[15-16] Hot electrons can then transfer transiently to adsorbate states by incoherent transport. This causes a sudden jump of the adsorbate to an excited potential energy surface (PES) that can lead to chemical transformation of the adsorbate or its desorption, a process referred to a desorption induced by electronic transitions (DIET).[17-18] CID, by contrast, represents a one-step interfacial charge transfer by coherent coupling of acceptor (adsorbate) and donor (plasmon) states. As such, CID can only take place when an adsorbate is present, while in the three-step process, hot electrons are created independent of the presence of an adsorbate. Another distinction between the two processes is their timescale. CID causes the decay of the SPR, through an electric dipole interaction, therefore it is an ultrafast process ($\sim$10-50 fs).[19-21] In contrast, DIET-like processes takes place after the SPR has already decayed by generating hot carriers. The partitioning between CID and DIET is determined by the coupling (or hybridization) of molecular and electronic metal states and the energy gap between the Fermi level and empty available energy states. If the coupling is strong and the plasmon energy is sufficient to populate the adsorbate states, then CID dominates the charge transfer to adsorbed molecules. Otherwise, DIET-like processes provides a more efficient route for metal-molecule charge transfer. In both cases, after a time determined by the lifetime of the adsorbate state, reverse charge transfer can take place from adsorbate to metal, which can leave the adsorbate in a vibrationally excited ground state.

Despite the growing acceptance of this framework, several open questions remain in our overall understanding of adsorbate-induced plasmon damping. One key question is how the electronic structure—specifically, the density of states—of adsorbed molecules influences the CID rate. Moreover, the mechanism of CID is still debated, involving not only a resonant electron transfer to resonance adsorbate states, but also the influence of molecular electric dipole moment[22] and (non-resonant) inelastic electron scattering.[23]

Previous studies have suggested that both changes in the DC resistivity of thin metal films and CID can be interpreted within a unified theoretical framework—specifically, diffuse electron scattering at the metal–adsorbate interface (see Supporting Information for details).[24] In the context of DC resistivity, this model is built around the concept of diffuse electron – surface scattering. In the absence of adsorbates, the DC surface resistivity is primarily governed by electron–phonon scattering.[25-26] Electrons that encounter a clean metal surface undergo predominantly specular (elastic) scattering due to the conservation of in-plane momentum—assuming minimal surface roughness or defects. However, when molecules are adsorbed onto the surface, this translational symmetry is disrupted. Electrons can then undergo diffuse (inelastic) scattering, in which they transiently transfer energy and momentum to the adsorbate by coupling into unoccupied molecular states.[27-29] This scattering behavior introduces additional resistance and is thus reflected in increased DC resistivity. Importantly, the electron–adsorbate scattering cross-section depends on the availability and energy alignment of adsorbate resonance states—*i.e.*, the density of states near the Fermi level since only electrons at the Fermi level carry current.

Experimental validation of this mechanism has been demonstrated through broadband IR reflectance measurements on copper films with adsorbed CO and $C_2H_4$.[27, 30] In these studies, increased diffuse electron scattering led to enhanced energy dissipation and a corresponding decrease in reflectance. This decrease in reflectance was found to scale with the observed increase in DC resistivity.[31] A similar mechanism—electron scattering at the metal–adsorbate interface—is also believed to underlie CID, but at optical (plasmon) frequencies.[32] While this idea is theoretically compelling, to our knowledge, there has been no direct experimental confirmation to date.

In this study, we explore how four different adsorbed molecules influence the DC electrical resistivity of gold (Au) thin films and their corresponding CID rates in plasmonic Au waveguides. By correlating electrical resistivity changes with CID rates, we identify two distinct regimes of plasmon-driven charge transfer. For adsorbates with resonance states overlapping the Au Fermi level, plasmon decay occurs via inelastic electron-adsorbate scattering. This process parallels DC resistivity changes and shows weak dependence on SPR energy. In contrast, for adsorbates with LUMO levels located above the Fermi level and within the plasmon energy range ($E_F + \hbar\omega_{SPR}$), direct metal-to-molecule electronic transitions dominate, exhibiting strong SPR energy dependence.

**Results**

The goal of this study is to investigate how changes in DC electrical resistivity correlate with the CID rates of four different molecular adsorbates on Au. By examining this relationship, we aim to gain deeper insight into the underlying mechanisms of metal–adsorbate energy and charge transfer.

We selected four test molecules as adsorbates: adenine (Ade), 4-aminothiophenol (ATP), biphenyl thiol (BPT), and 1-dodecanethiol (DDT) (Figure 1a). For electrical resistivity measurements, the molecules were adsorbed onto 30 nm-thick Au films, and the surface resistivity change was measured through a 4-point-probe setup (schematically shown in Figure 1b). CID rates were determined by measuring plasmon propagation losses induced by each molecule on Au plasmonic waveguides (Figures 1c–d). We assume that the adsorption behavior is similar on thin films and waveguides, particularly for the thiol-containing molecules (ATP, BPT, DDT), which form strong covalent bonds with Au surfaces. This assumption is likely valid given the well-known surface binding characteristics of thiols.[33]

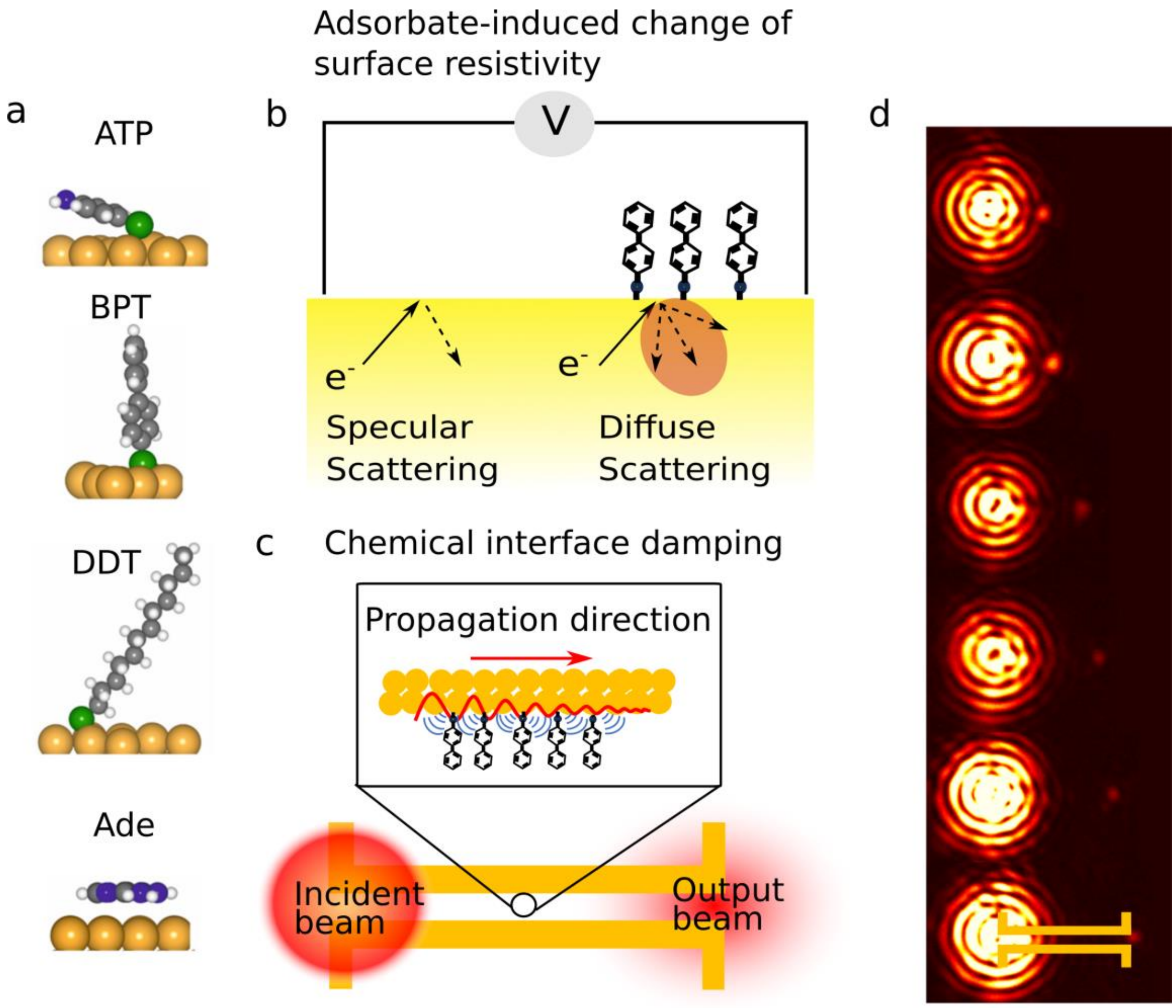

**Figure 1. Scheme of the molecular systems used and experimental setup.** a) The calculated adsorption geometry of 4-aminothiophenol (ATP), bi-phenyl thiol (BPT), dodecanethiol (DDT) and adenine (Ade) on gold. b) Scheme of adsorbate-induced change of surface resistivity through diffuse electron-adsorbate scattering. c) Scheme of the adsorbate-induced plasmon propagation loss in Au waveguides. d) An optical image of the excitation and output beams in waveguides ranging from 1.5 to 4 µm length. A sketch of the waveguide structure is shown on the right-most waveguide. The bright lower spots are scattered light from the input beam. The upper spots are light scattering from the end of each waveguide of varying length.

**Adsorbate-induced change in resistivity.** Experimentally, the surface resistivity of the Au films is measured by using a collinear 4-point probe. This method allows the measurement of the film resistivity with high accuracy. The metal film resistivity can then be determined as $\rho = R_{sheet} \times d$ , where $R_{sheet}$ is the sheet resistivity and $d$ is the metal film thickness. Figure S2 a) shows the resistivity of blank Au films of different thicknesses. As expected, as the Au film thickness decreases below the mean free path length ($\sim 40\ nm$ at RT), the resistivity of the Au film increases due to increased diffuse electron surface scattering from surface defects. In Figure S1 b) and c) we show the relative change of resistivity induced by the four different adsorbed molecules for a 30 and 15 nm thick Au film.

The adsorbate-specific DC scattering cross-section is related to the initial slope of the resistivity increase versus the number of adsorbates, $\partial\rho/\partial n_a$:[34-35]

$$\Sigma_{DC} = \frac{16ne^2 d}{3mv_F}\frac{\partial\rho}{\partial n_a}\bigg|_{n_{a\to 0}}$$

Where $n$ is the electron concentration density of the metal, $d$ is the metal thickness, $m$ is the electron mass and $v_F$ is the Fermi velocity.

Figure 2 a) shows the initial time-dependent change of the Au resistivity (30 nm thickness) upon the adsorption of the four different adsorbates. At time 0, the respective molecular solutions (1 mM each) were added to the Au film, starting the molecular adsorption process. The DC resistivity follows a typical behavior: it increases rapidly in the first seconds due to molecular adsorption and the formation of a monolayer and then it reaches a plateau. Upon longer exposure overnight (*i.e.*, Figure S1 b)-c)) the molecular packing increase and molecular re-orientation due to inter-molecular van der Waals forces cause a gradual increase in resistivity.[36]

We determined the adsorbate-induced change in resistivity for one adsorbed monolayer. The time necessary for the formation of a monolayer was taken as the point where the slope of the resistivity curve begins to change. For all the thiolated molecules (ATP, BPT, DDT) the monolayer formation is very fast, within the first 2 seconds, due to the covalent Au-S bond.[33] For Adenine, the monolayer formation is slower, so the first 10 seconds were used to determine the resistivity change, after which we observed a plateau in the relative change of resistivity (Figure 2a). The adsorption behavior of all four molecules on Au has been extensively analyzed in the past, therefore the surface density of adsorbed molecules per monolayer is known (see Table 1). Consequently, we could determine the DC diffuse scattering cross-section $\Sigma_{DC}$ for each adsorbed molecule (Table 1). The scattering cross-sections obtained by us are similar to previous values for chemisorbed molecules and metal thicknesses (for example $\sim 19$ Å$^2$ for O on Cu).[34]

**Table 1. Adsorbate-induced change in resistivity.** The experimentally determined values of the adsorbate-induced change in resistivity per adsorbed monolayer, the number of adsorbed molecules per monolayer and the obtained DC electron scattering cross section.

| | $\frac{\Delta\rho}{ML}[n\Omega.m]$ | $^{n_a}/_{ML}\,[nm^{-2}]$ | $\Sigma_{DC}[\text{Å}^2]$ |
|---|---|---|---|
| ATP | 6.6±2.3 | 4[37] | 31±10 |
| BPT | 4.64±0.7 | 4[37] | 22±3.33 |
| DDT | 5±1 | 5[38] | 19±4.15 |
| Ade | 0.28±0.1 | 1.5[39] | 3.58±1.25 |

The scattering cross-section of Adenine agrees well with the theoretical prediction based on the calculated DOS within the error bars (Figure 2b). However, for BPT, ATP and DDT a higher scattering cross-section was observed experimentally (Figure 2b). The DFT calculations showed that adsorbed BPT, ATP and DDT have high perpendicular dipole moments of -5.6, -3.1 and -2.85 D, respectively, while Adenine has a much smaller perpendicular dipole moment (Tables S3, S4). Therefore, we attribute the additional scattering cross-section for BPT, ATP and DDT to the influence of the perpendicular dipole moment.[40] This is consistent with BPT showing the highest deviation from the expected scattering cross-section, while ATP and DDT have similar perpendicular dipole moments, and show similar deviations. The total DC diffuse scattering cross section is the sum of the perpendicular dipole moment contribution ($\Sigma_{\mu}$) and the molecular density of states contribution ($\Sigma_{DOS}$). For $\Sigma_{\mu}$ , we infer a value of $\sim 2.5 - 3$ Å$^2$ per Debye for the diffuse scattering cross-section due to the perpendicular dipole moment, which may be a useful number for future comparison and reproduction of these results. The influence of the perpendicular dipole moment on electron transport in molecular junction devices has been well documented.[41-44] Intuitively, a larger perpendicular dipole moment represents a larger perturbation on scattering electrons. In addition, a larger perpendicular dipole moment of the adsorbed molecule will result in a larger local work function shift, effectively decreasing the barrier for electron tunnelling to molecular states (*i.e.*, diffuse scattering). In our case, the metal electrons undergo a higher scattering rate due to the lower metal work function than would be expected based on the molecular density of states at the metal Fermi level alone, shown in Figure 2c.

**Chemical interface damping.** Our aim was to compare the DC diffuse electron scattering cross-sections obtained from the change of resistivity to the CID rate of the same adsorbed molecules to verify if they are related to the same electron scattering phenomenon. CID can be measured from the loss of SPPs at metal-dielectric interfaces as it is attributed to electron scattering and depends on the degree of confinement, the quality of fabrication and the surface characteristics of the metal. We define a loss coefficient, $\alpha = \alpha_b + \alpha_{rad} + \alpha_{surf} + \alpha_{CID}$, where $\alpha_b$ describes nonradiative scattering in the bulk of the metal and $\alpha_{rad}$ describes radiative damping. Surface damping can be separated into two contributions: electron scattering at the metal surface, $\alpha_{surf}$ and CID, $\alpha_{CID}$, induced by adsorbates on the metal surface. The first three loss mechanisms can be viewed as intrinsic to the plasmonic waveguide, which we denote with an intrinsic loss coefficient, $\alpha_i = \alpha_b + \alpha_{rad} + \alpha_{surf}$.

The effect of CID at optical frequencies can be studied by measuring the propagation loss of SPP waves. Here, we have studied CID in strongly confined gap plasmon waveguides. Light at a wavelength of 790 nm is confined in a <100 nm wide gap between two metal wires, as shown in Figure 1c, d. Light is coupled to the waveguides via antenna couplers with a free space beam to SPP wave coupling efficiency of about 30% (see Table S2, SI).[45] As shown in Figure 1d, the CW laser was tightly focused (diffraction-limited) to one end of the waveguide via an oil-immersion objective lens with a high numerical aperture (NA = 1.45). The polarization of the laser was set parallel to the antenna to enable efficient in-coupling to SPP waves that propagate along the waveguide and then couple out via the antenna at the opposite end.

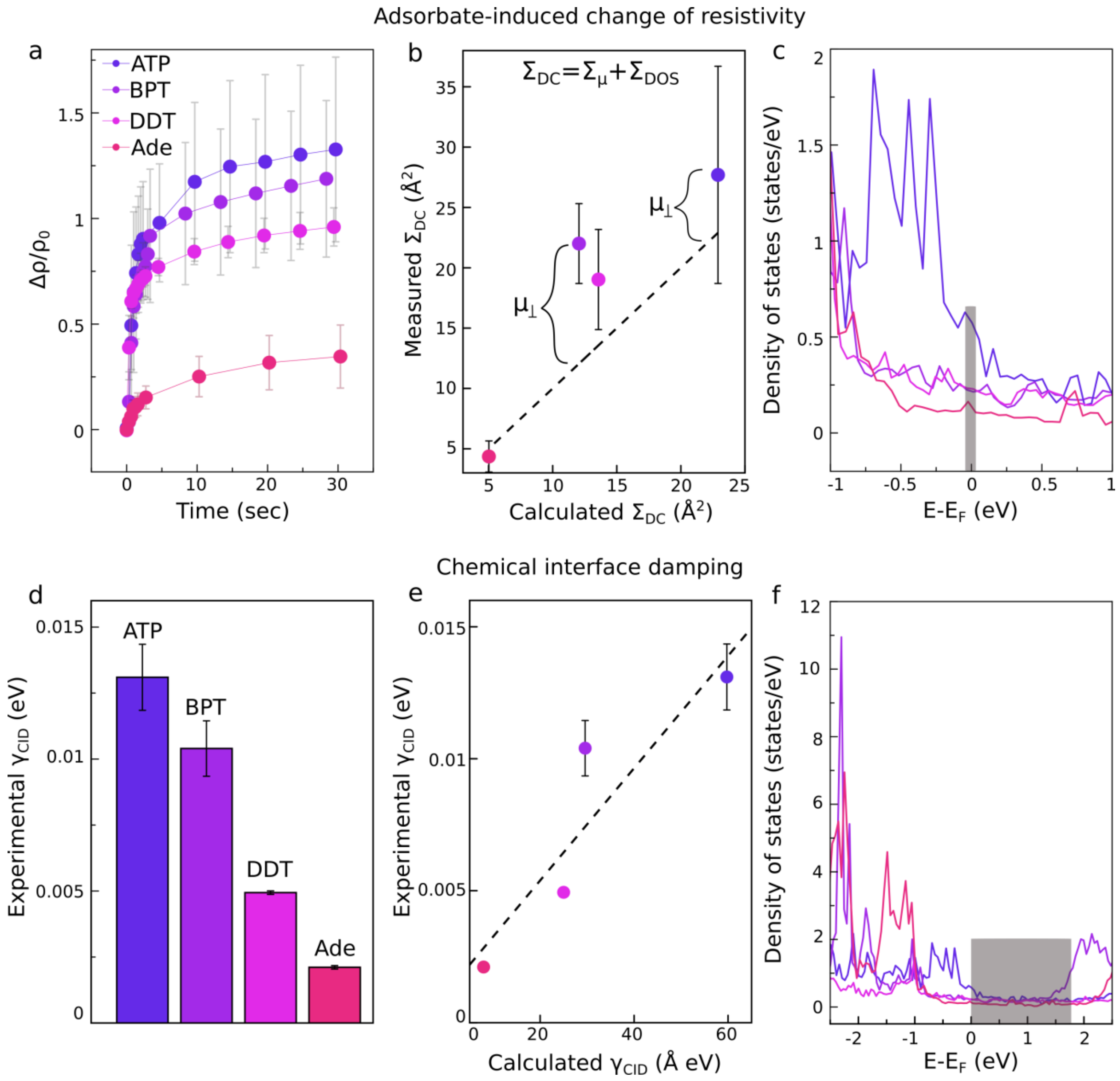

**Figure 2. Theoretical and experimental values for adsorbate-changes of DC resistivity and optical chemical interface damping rate.** a) Time-dependent surface resistivity measurements of Au films (30 nm thickness) with adsorbed ATP, BPT, DDT and Ade. The slope of the increase in resistivity was determined from the linear part of the curve: the first 2 seconds were considered for ATP, BPT and DDT and the first 10 seconds for Ade. b) The calculated and experimentally determined DC electron scattering cross-section for each adsorbed molecule. The dashed line represents the calculated scattering cross-section considering only the molecule density of states, while $\mu_\perp$ represents the contribution from the perpendicular dipole moment of the adsorbed molecule. c) The density of states of the adsorbed molecule. Only the density of states at the Fermi level energy (grey rectangle) contributes to the adsorbate-change of DC resistivity. d) The experimental CID rate, $\gamma_{CID}$. e) Comparison of the experimental and calculated $\gamma_{CID}$. The

dashed line was obtained by linear fitting. Since the effective length of the Au waveguides is not well defined, the calculated values of $\gamma_{CID}$ are expressed as $\text{Å} \cdot eV$ (see the text for details). f) The density of states of the adsorbed molecules. The shaded area represents the energy span of the plasmon resonance.

The CID effect was studied by comparing the change in loss coefficient of plasmonic waveguides with ($\alpha$) and without ($\alpha_i$) molecular adsorbates. This method was recently used to quantify CID on Au nano-stripes.[46-47] SPP loss coefficients were measured along gap plasmon waveguides using the cut-back method,[48] where waveguides with different lengths ranging from 1.5 µm to 4.0 µm were measured. The transmission of light, $T(L)$, through the waveguides of varying length, $L$, was measured and fit with a model accounting for the average loss coefficient and the in/out waveguide coupling efficiency, $\eta$, given by $T(L) = \eta^2 e^{-\alpha L}$ . The intrinsic loss, $\alpha_i$, of a selection of plasmonic waveguides of varying length were first measured without any adsorbed molecules to acquire statistically significant data, after which the same plasmonic waveguides were functionalized with the four different molecules and the transmission efficiency was measured again. The average transmission, $T(L)$, of waveguides with lengths ranging from $L = [1.5, 4]$ µm was fitted to the above equation and plotted in Figure S3. Next, the CID rate was determined by: $\gamma_{cid} = \frac{c*\alpha_{cid}}{2.58}$, where 2.58 is the propagating plasmon group velocity.[45]

The extracted $\gamma_{cid}$ are shown in Figure 2d and the coupling efficiencies and loss coefficients for the various molecular adsorbate configurations are shown in Table S2. Next, we compared the experimental $\gamma_{cid}$ to the calculated values by using the theoretical method introduced by Persson (see the SI for the calculation details).[32] Usually, CID is determined using metal nanoparticles with a fixed radius or effective size. However, in our case, waveguides with different lengths were used for determining the plasmon propagation loss. Therefore, it was not possible to calculate the CID rate in eV; instead, we calculated $\gamma_{CID}\ R\ [eV\ \text{Å}]$. Since we used the same waveguide size for all four molecules, this is a good measure of the relative difference between the CID rates of the different molecules.

Of the four molecules, ATP has the strongest effect on the plasmon propagation length, and on DC resistivity. This can be explained by the fact that ATP has the highest occupied molecular orbital (HOMO) overlapping the metal Fermi level energy, supporting thus the electron scattering model (see Figure 2c, f). For both adsorbate-change of resistivity and CID rates, we observed a similar trend among the four adsorbates. A good agreement between the experimental and calculated results was observed, given the relatively simple model and the approximations that were made. For example, $\gamma^{exp}_{CID,ATP} \big/ \gamma^{exp}_{CID,Ade} \approx 7.3$ while the calculated CID rates of ATP and Ade is $\approx 6.5$, giving a relative discrepancy of approx. 10%. BPT shows the biggest difference between the calculated and measured CID rate, which will be analyzed in the following sections.

In Figure 3, we compare directly the DC electron scattering cross-section and the CID rates. In the case of the DC electron scattering cross-section, we used the value of $\Sigma_{DOS} = \Sigma_{DC} - \Sigma_{\mu}$, which considers only the effect of the molecular DOS near the Fermi energy (Figure 3b). This is justified, since in principle $\Sigma_{DOS}$ and $\Sigma_{\mu}$ act independently[40] and the perpendicular molecular dipole moment does not influence the CID rates,[22] allowing us to subtract its effect when comparing the two phenomena. A good correlation between $\Sigma_{DOS}$ and the CID rate is observed both for the theoretical and the experimental data. For example, $\Sigma^{exp}_{DC,ATP}/\Sigma^{exp}_{DC,Ade} \approx 6.5$ which is nearly identical to the CID rate of the two adsorbed molecules: $\gamma^{exp}_{CID,ATP}/\gamma^{exp}_{CID,Ade} \approx 6.5$.

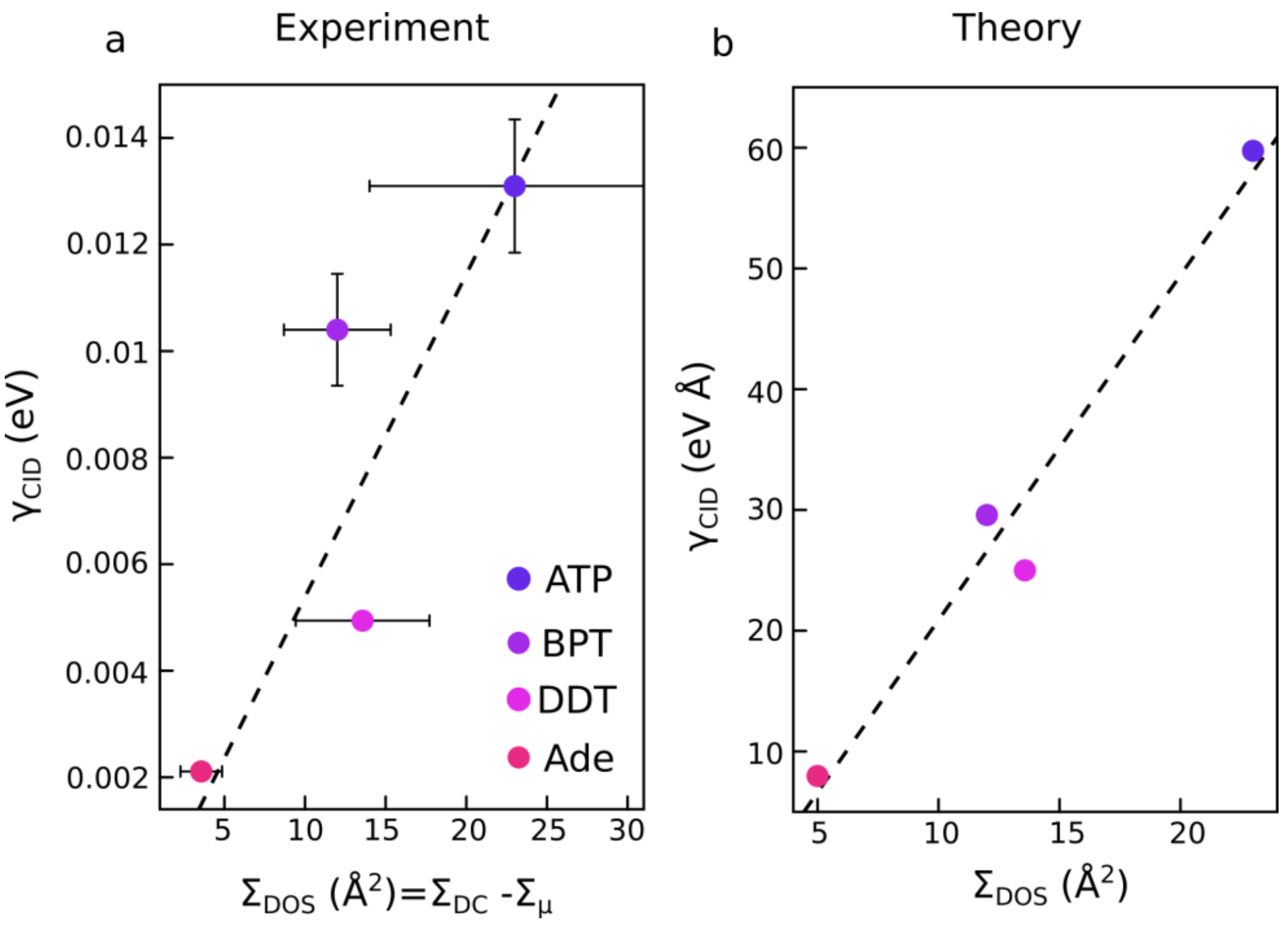

**Figure 3. Comparison of the DC electron scattering cross-section and the chemical interface damping rate.** a) Comparison of the experimentally determined $\gamma_{CID}$ and the DC scattering cross-section, $\Sigma_{DOS}$ (see text). b) Comparison of the calculated $\gamma_{CID}$ and $\Sigma_{DOS}$. In all plots the dashed line represents the linear fit.

The biggest difference between the experimental and theoretical CID rates is for BPT (see Figure 2e and Figure 3a). The experimental value obtained for the CID rate of BPT is consistently higher than the calculated value obtained theoretically, and it also differs the most from its DC scattering cross-section. To understand this discrepancy, wavelength-dependent CID measurements at 790 and 860 nm were performed for ATP and BPT (Figure 4). For ATP there is no significant difference in the CID rate at 790 and 860 nm within the error rate, whereas for BPT there is a ~30% decrease in the CID rate at 860 nm compared to 790 nm (Figure 4 a, b). This decrease in the CID rate is due to the LUMO orbital BPT being centered at ~2.1 eV above the Fermi level (Figure 4 c). At 790 nm (1.57 eV), the plasmon energy can excite direct electronic transitions to this adsorbate state whereas at 860 nm (1.44 eV) this direct charge transfer pathway is less efficient.

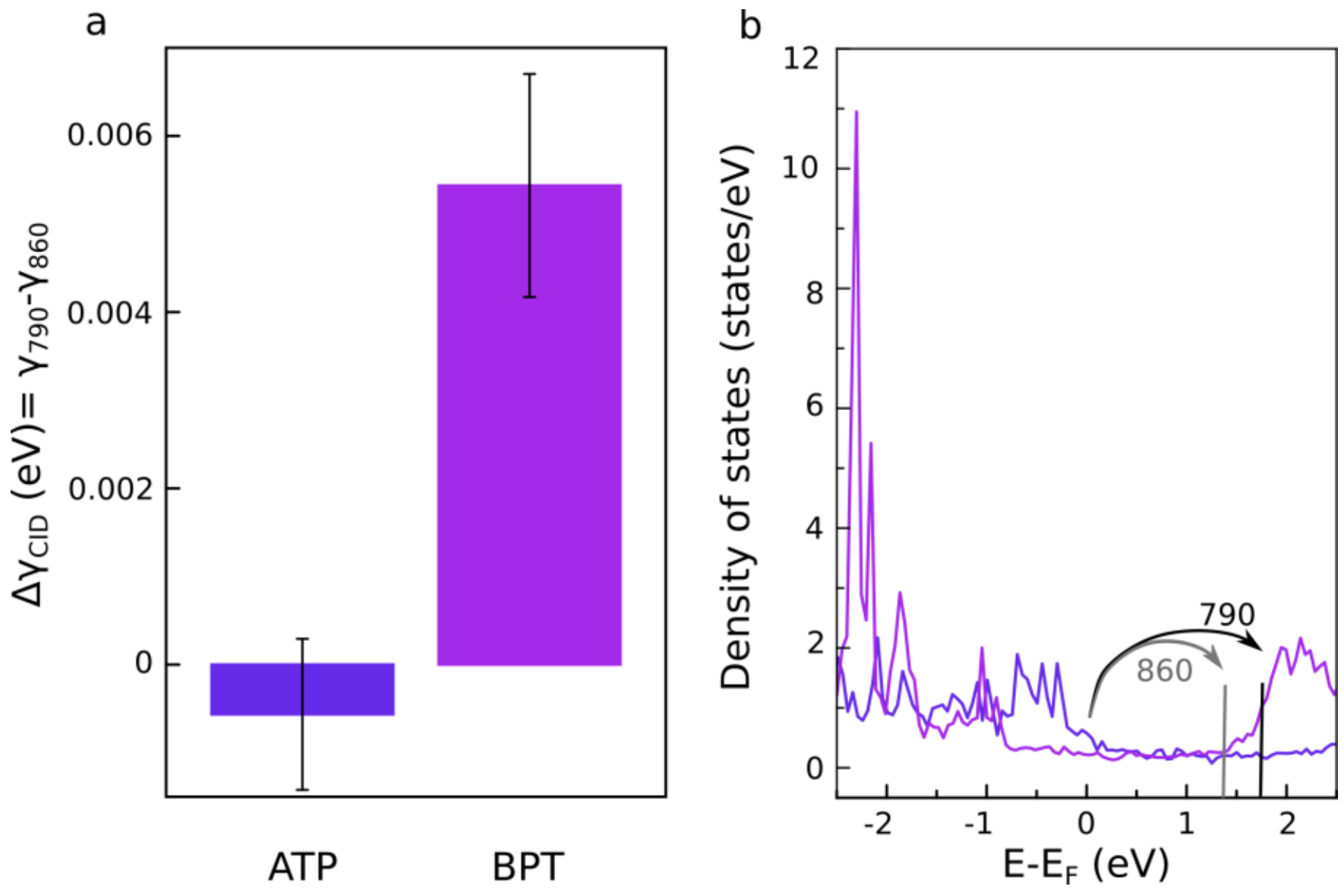

**Figure 4. Wavelength-dependent chemical interface damping rate.** The measured differential chemical interface damping rate at 790 and 860 nm for ATP and BPT. b) The density of states of adsorbed ATP and BPT. The arrows indicate the possible electron transfer transitions at 790 (1.57 eV) and 860 nm (1.44 eV), respectively.

## Discussion

These results highlight two regimes by which adsorbed molecules can impact SPPs decay (*i.e.*, CID), depending on the density of states of the metal – adsorbate system and the plasmon energy (Figure 5). (i) The first regime of CID is the direct, one-step charge transfer from the plasmon state to the adsorbate resonance state. It was shown before that this regime involves coherent charge transfer between interfacial, strongly coupled (hybridized) metal-adsorbate states which are resonant with the plasmon energy.[21, 49-50] For the adsorbate systems studied here, this mechanism is strongest for BPT due to its LUMO orbital being resonant at energies within the plasmon energy. This is most evident in the wavelength-dependent CID rate of BPT (Figure 4a) as well as the fact that the CID rate of BPT deviates the most from its DC scattering cross-section, which only probes the density of states at the Fermi level energy (Figure 3).

(ii) The second regime of CID is the analogue of adsorbate-induced change of DC resistivity by electron scattering at the metal-molecule interface. This effect is distinct from the commonly assumed CID mechanism (*i.e.*, regime (i) described above) in which the plasmon energy is resonant with the energy difference between the metal Fermi level and an empty resonance state. In the regime described here, adsorbate-induced plasmon decay takes place through coupling between SPPs and adsorbate degrees of freedom (such as nonadiabatic vibrational coupling), without the need for a resonant transition.[51-53] Such interactions are well-known for example from vibrational damping of adsorbates through electronic friction.[54-56] This regime dominates in cases where the adsorbate empty states are at higher energies than the plasmon energy compared to the Fermi level[23] and the CID rate is proportional to the density of states at the Fermi level, as revealed by comparing CID and DC surface resistivity measurements. In the systems tested in this study, this regime is most obvious for ATP, which has its HOMO level overlapping the Fermi level. This is supported by the good qualitative correlation between the DC scattering cross-section and the CID rates for the four adsorbates (Figure 3) as well as by the fact that the CID rate of ATP does not change when decreasing the SPP energy (Figure 4 a). As was emphasized recently, plasmon decay is a purely quantum effect and cannot be interpreted as a gradual loss through a "frictional force".[15] In this study we used the semiclassical analogy of a frictional force between the electron current and adsorbates to gain a more intuitive view of the phenomena (as done in the original article describing the adsorbate-induced change in DC resistivity through electron scattering[34]), but it should be noted that this is a quantum mechanical effect. A fully quantum mechanical model for adsorbate-change of resistivity was developed later.[57]

These two regimes of adsorbate-induced plasmon damping were recently studied with time-resolved spectroscopy.[23] One might draw a parallel between the two CID regimes and fluorescence and non-resonant Raman scattering. For fluorescence, the molecule is resonantly excited to the first excited electronic level, similar to the first CID regime described above in which the molecule is promoted to an excited PES. For non-resonant Raman scattering, intra-molecular vibrational transitions are excited non-resonantly through a range of intermediate energy levels, like in the second CID regime described above in which multiple low-energy electron-hole pairs couple to molecular degrees of freedom.

We used the most common theoretical model for CID, developed by Persson,[32] to model the CID induced by the four adsorbates (see the first section of the SI). The theoretical model relies on the adsorbate-induced damping of the parallel, $\gamma_{\parallel}$, and perpendicular, $\gamma_{\perp}$, electric fields. Note that, even though at visible wavelengths the electric near field outside the metal surface is mostly perpendicular, inside the metal surface the perpendicular electric field is screened very effectively while the parallel electric field interacts stronger with the metal electrons. Intuitively, $\gamma_{\perp}$ would describe the first regime, coherent electron transfer to strongly coupled (hybridized) adsorbate resonant states through electric dipole interactions and be dominant for BPT, while $\gamma_{\parallel}$ describes the second regime, inelastic electron scattering at the metal-molecule interface and be dominant for ATP. In fact, $\gamma_{\parallel}$ is characterized by a scattering cross-section, which, in the limit of $\omega_{SPP} \rightarrow 0$, yields directly the DC scattering cross-section providing the link between CID and adsorbate-induced surface resistivity, as previously suggested.[24, 32] By analyzing the plasmon-energy dependent $\gamma_{\parallel}(\hbar\omega_{SPP})$ it is clear that it captures qualitatively well the experimental results, which is best observed in the extreme cases of ATP and BPT (Figure S1). As the plasmon energy approaches 0 we recover the DC scattering cross-section, with ATP having the biggest cross-section. At plasmon energies between $\sim 0.7 - 2\ eV$, $\gamma_{\parallel}(\hbar\omega_{SPP})$ reaches a plateau for ATP, which explains the wavelength-dependent CID results (Figure 4). At plasmon energies higher than $\sim 2\ eV$, $\gamma_{\parallel}(\hbar\omega_{SPP})$ of BPT becomes dominant over ATP due to its LUMO state centered at $\sim 2\ eV$.

On the other hand, the perpendicular component, $\gamma_{\perp}$, which intuitively would characterize the direct electron transfer between strongly hybridized metal-adsorbate states, is more difficult to interpret physically. The plasmon-energy dependent perpendicular component, $\gamma_{\perp}(\hbar\omega_{SPP})$, exhibits fewer features showing only an exponential increase as the plasmon energy approaches the LUMO state of the adsorbates (Figure S1). However, note that $\gamma_{\perp}(\hbar\omega_{SPP})$ for ATP is larger than BPT even for plasmon energies $\sim 1.5 - 2\ eV$ which should be enough to resonantly excite electrons from the Fermi level to the LUMO orbital of BPT. This is the likely reason why the theoretical value for the CID rate of BPT is smaller than the experimental one and deviates the most out of all four adsorbates tested (see Figure 2 e and 3). We believe that the theoretical model for CID should be improved in order to capture more accurately these two regimes of adsorbate-induced plasmon damping which depend on the density of states of the metal-adsorbate system and the plasmon energy. In related fields, such as photosynthetic light harvesting and energy transfer[58], there is still an effort to construct such physical models which can describe both coherent and incoherent photoexcited electron transfer.[59-61]

Another mechanism of adsorbate-induced plasmon damping that was suggested previously involves plasmon scattering from the adsorbate in-plane dipole moment.[22] In our systems we could not unambiguously quantify the influence of the in-plane dipole moment of adsorbates to CID. ATP has the highest in-plane dipole moment and the highest density of states close to the Fermi level, therefore it is difficult to disentangle the two effects. We believe that, if there are molecular states close to the Fermi level, within the plasmon energy, this pathway will dominate over plasmon decay through adsorbate dipole moment.

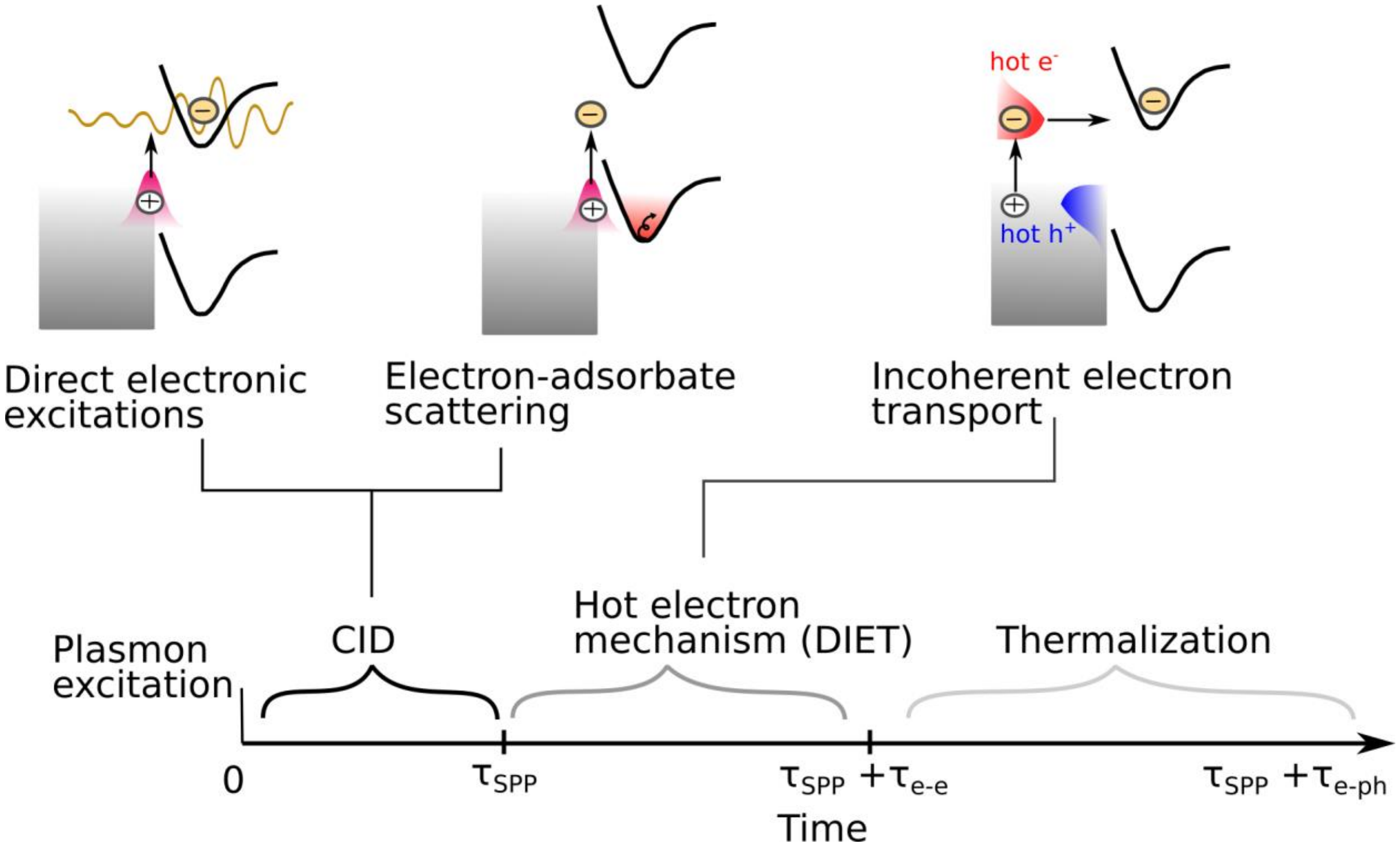

**Figure 5. Mechanisms of charge plasmon-driven charge transfer.** Following plasmon excitation, within the plasmon decay time, $\tau_{SPP}$, there can be a direct (coherent) electronic excitation of adsorbates or electron-adsorbate scattering, depending on the position of the adsorbate resonance states. After the plasmon decayed into hot carriers, e-e scattering takes place at the same time with possible incoherent transport of hot carriers into adsorbate states. During this time the metal lattice temperature does not change. Finally, at times longer than $\tau_{SPP} + \tau_{e-e}$, the electron energy is transferred back to the metal lattice.

## Conclusions

In summary, we investigated the adsorbate-induced changes in both DC electrical resistivity and plasmon damping (CID) for four different molecular adsorbates on Au surfaces. Our results reveal the existence of two distinct regimes of CID, depending on the metal-adsorbate electronic structure (*i.e.*, density of states) and plasmon energy (Figure 5).

In the first regime, exemplified by BPT – which has its LUMO level within the plasmon energy – plasmon damping occurs through direct electronic transitions between strongly coupled (hybridized) metal-adsorbate states. Since this effect involves a resonant transition, it is strongly dependent on plasmon energy. This process is analogous to the mechanism of direct electronic excitations known from femtosecond laser-driven desorption studies.[62]

In the second regime, exemplified by ATP—which has a resonance state overlapping the Fermi level and the LUMO level at higher energy above the Fermi level—plasmon damping is dominated by inelastic electron scattering, wherein low-energy e–h pairs play a central role. In this regime, adsorbate-induced plasmon damping occurs via plasmon coupling to molecular degrees of freedom, such as nonadiabatic vibrational coupling, potentially leading to ultrafast vibrational heating.[23] This damping mechanism is weakly dependent on the plasmon energy, suggesting that plasmon damping occurs through electron–adsorbate scattering rather than through direct electronic transitions to discrete molecular orbitals. Similar behavior was observed for Ade and DDT, both of which have the LUMO at higher energies above the Fermi level compared to the plasmon energy. Notably, the same electron–adsorbate scattering process responsible also contributes to adsorbate-induced changes in DC resistivity. As such, measurements of resistivity may serve as a valuable indirect probe of this energy transfer channel.

**Materials and methods.**

**Plasmonic waveguide fabrication.** The plasmonic waveguides were fabricated via electron-beam lithography, on 170 µm thick borosilicate substrates. The substrate was submerged in acetone, sonicated, rinsed with isopropyl alcohol (IPA), dried with compressed nitrogen and plasma ashed with oxygen for 5 min. A poly(methyl methacrylate) resist, PMMA 950 A4, was spin-coated onto the substrate (4,000 revolutions per min, 1 min) and baked (180 ºC, 18 min). To prevent charge build-up, a layer of the conductive polymer E-spacer was spin-coated on top of the PMMA (2,000 revolutions per min, 1 min) and baked (90 °C, 30 sec). An electron lithography system (Raith) was used to expose the polymer in a predefined pattern at 20 kV with a 10 µm aperture. The patterned sample was rinsed in DI water to remove the conductive polymer and developed in a solution of methyl isobutyl ketone (MIBK) and IPA (3:1) for 30 s, followed by 30 s in IPA to halt development. After rinsing in IPA and drying with compressed nitrogen, Au (50 nm) was deposited on to the developed polymer using a thermal evaporator (Angstrom), with an initial Cr layer (2 nm) to adhere Au to the substrate. To remove the polymer and excess metal, the sample was submerged in acetone for 24 h, rinsed with IPA and dried with compressed nitrogen. Further oxygen plasma ashing removes residual polymer, which was confirmed with Raman spectroscopy of the fabricated waveguides.

**DC surface resistivity measurement.** For surface resistivity measurements, Au films of various thicknesses were deposited through electron-beam deposition at a rate of approx. 1 Å/s on regular cover glasses (borosilicate glass, 24 x 24 mm, 0.13-0.17 mm thickness). A Cr adhesion layer of 2 nm was deposited by electron-beam deposition prior to Au deposition to ensure the stability of the Au film. The surface resistivity was measured using a collinear 4 point-probe setup (Ossila Ltd.) with a probe spacing of 1.27 mm. 25 measurements were acquired for each measurement and the average is shown.

**DFT simulations.** To obtain the adsorbate contribution to the density of states (DOS) which enters Persson's model for CID, ab initio DFT calculations were performed using a periodic slab model for the Au surface with various adsorbates, see SI for details. First, the atomic structure was determined using geometry optimization. Then, the electronic DOS projected onto the adsorbate orbitals was obtained.

To calculate the dipole moment induced in the Au surface by the adsorbates, DFT calculations were performed using a cluster model of the surface with a single adsorbed molecule. The induced dipole was obtained by subtracting the dipole of the cluster without adsorbate and the dipole of the molecule in the gas phase from the dipole of the cluster with adsorbate, see SI for details.

**Acknowledgements.**

P.N. and N.J.H. acknowledge support from the Robert A. Welch Foundation under grant numbers C-1222 and C-1220. N.J.H., P.N. and E.C. acknowledge the Institute for Advanced Study (IAS) from Technische Universität München (TUM) for financing the focus group on 'Sustainable photocatalysis using plasmons and 2D materials (SusPhuP2M)' as part of the Hans Fisher Senior Fellowships program. We acknowledge funding and support from the Deutsche Forschungsgemeinschaft (DFG) under Germany´s Excellence Strategy – EXC 2089/1 – 390776260 e-conversion, the Bavarian program Solar Technologies Go Hybrid (SolTech) and the Center for NanoScience (CeNS). A.S. acknowledges support from the Alexander von Humboldt foundation.

Supporting Information for

# Optical and electrical probing of plasmonic metal-molecule interactions

Andrei Stefancu*[1], Wenxuan Tang*[2], Ming Fu[2], Jordan Edwards[3], Naomi J. Halas[4], Ross C. Schofield[2], Toby S. Millard[2], Peter Nordlander[4], Johannes Lischner[3], Pilar Carro[5], Rupert Oulton*[2], Emiliano Cortes*[1]

1 Nanoinstitute Munich, Faculty of Physics, Ludwig-Maximilians-Universität (LMU), Munich, Germany

2 Blackett Laboratory, Imperial College London, London, UK

3 Department of Materials, Imperial College London, South Kensington Campus, LondonSW7 2AZ, United Kingdom

4 Department of Electrical and Computer Engineering, Department of Chemistry, Department of Physics and Astronomy, Department of Material Science and Nanoengineering, and Laboratory for Nanophotonics, Rice University, Houston, Texas 77005, United States, Technical University of Munich (TUM) Institute for Advanced Study (IAS), Garching, Germany

5 Department of Chemistry, Faculty of Sciences, University of Laguna, Institute of Materials and Nanotechnology, 38200 La Laguna, Spain

**Description of the theoretical model**

The theoretical model of CID introduced by Persson considers the adsorbate damping of parallel, $\gamma_{\|}$, and perpendicular, $\gamma_{\perp}$ electric fields.

The parallel component, $\gamma_{\|}$, is characterized by an adsorbate-dependent electron scattering cross-section which, in the limit of plasmon frequency, $\hbar\omega_{SPP} \to 0$ yields the electron scattering cross-section for DC conditions (i.e., adsorbate-induced DC surface resistivity change). The adsorbate-dependent diffuse scattering cross-section can be computed as:[1-2]

$$\sigma_{diff}(\omega) = \sigma_0 J(\omega)$$

$$\sigma_0 = \frac{64\omega_F Q}{3\pi n v_F}$$

Where $\omega_F$ is the Fermi frequency, $n$ is the density of metal electrons, $v_F$ the Fermi velocity and $Q$ is a number depending on the symmetry of the molecular orbital (0.33 for p orbitals and 0.2 for s orbitals).

At the plasmon resonance frequency, for CID, the $J(\omega)$ integral is given by:

$$J(\omega_{SPR}) = \frac{\pi}{4\hbar^2\omega_F}\frac{1}{\omega_{SPR}}\int_{E_F-\hbar\omega_{SPR}}^{E_F} dE[E\Gamma\rho(E+\hbar\omega_{SPR}) + (E+\hbar\omega_{SPR})\Gamma\rho(E)]$$

While for DC excitation (*i.e.*, $\omega = 0$) this becomes:

$$J(0) = \frac{\pi}{2}\Gamma\rho(E_F)$$

From this, we can calculate the change in the plasmon decay due to the damping of the parallel surface electric fields as $\Delta\gamma_{\|}[\text{Å}\ eV]$ and the DC scattering cross-section, $\Sigma_{DC}$:

$$\Delta\gamma_{\|} = {}^{3}\!/_{8R}\, v_F n_a \sigma_{diff}(\omega_{SPR})$$

$$\Sigma_{DC} = \sigma_{diff}(0)$$

Likewise, the damping of normal surface electric fields, $\gamma_{\perp}$ due to adsorbates is:

$$\Delta\gamma_{\perp} = \frac{16\pi}{1+2\varepsilon_0} n_a \omega_{SPR} Im\alpha_{\perp}(\omega_{SPR})$$

Where:

$$Im\alpha_{\perp}(\omega_{SPR}) = 2\pi(ed)^2\int_{E_F-\hbar\omega_{SPR}}^{E_F} dE\ \rho(E)\rho(E+\hbar\omega_{SPR})$$

Where $d$ is the distance between the metal and the image plane of the molecular orbital.

The total CID rate can be written as:

$$\gamma_{CID} = \gamma_{\parallel} + \gamma_{\perp}$$

The total CID rate was calculated with a Python script adapted from ref. [3-4]

**Table S1.** Parameters used for calculating the CID and SR.

| Parameter | Description | Value |
|---|---|---|
| $n$ | Carrier concentration of core metal | 0.059 $Å^{-3}$ |
| $E_F$ | Fermi energy of core metal | 5.53 eV |
| $\Gamma$ | Linewidth of adsorbate induced resonance or virtual state | 1 eV |
| $\varepsilon_0$ | Bulk dielectric constant of the adsorbate | 2[5] |
| $Q$ | A number that mainly depends on the symmetry of the resonance state. | 0.33 (for BPT) and 0.2 (for all other adsorbates) |
| $d$ | Distance between dynamic image plane and center of mass of orbital | 0.9 Å [6] |
| $v_F$ | Fermi velocity of core metal | $1.4 \times 10^{-16}$ Å/s |

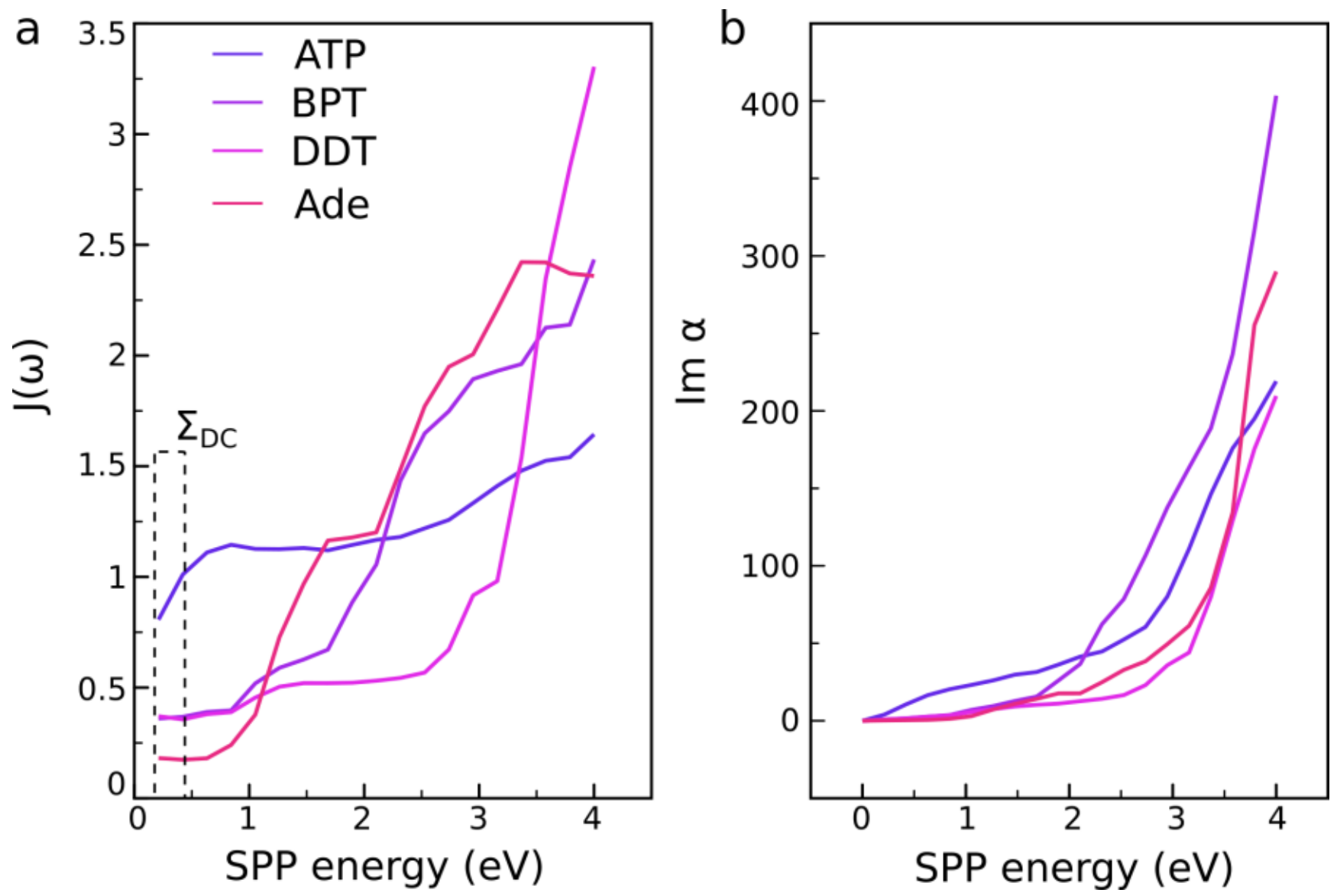

**Figure S1.** The plasmon energy dependent a) $J(\omega)$ and b) $Im\ \alpha$ integrals for the four adsorbates. Note that, even if the integrals in the case of Adenine are large, the CID rate depends on the adsorbate concentration, $n_a$, which is smallest for Adenine compared to the other adsorbed molecules tested here.

**Supplementary DC Surface Resistivity data**

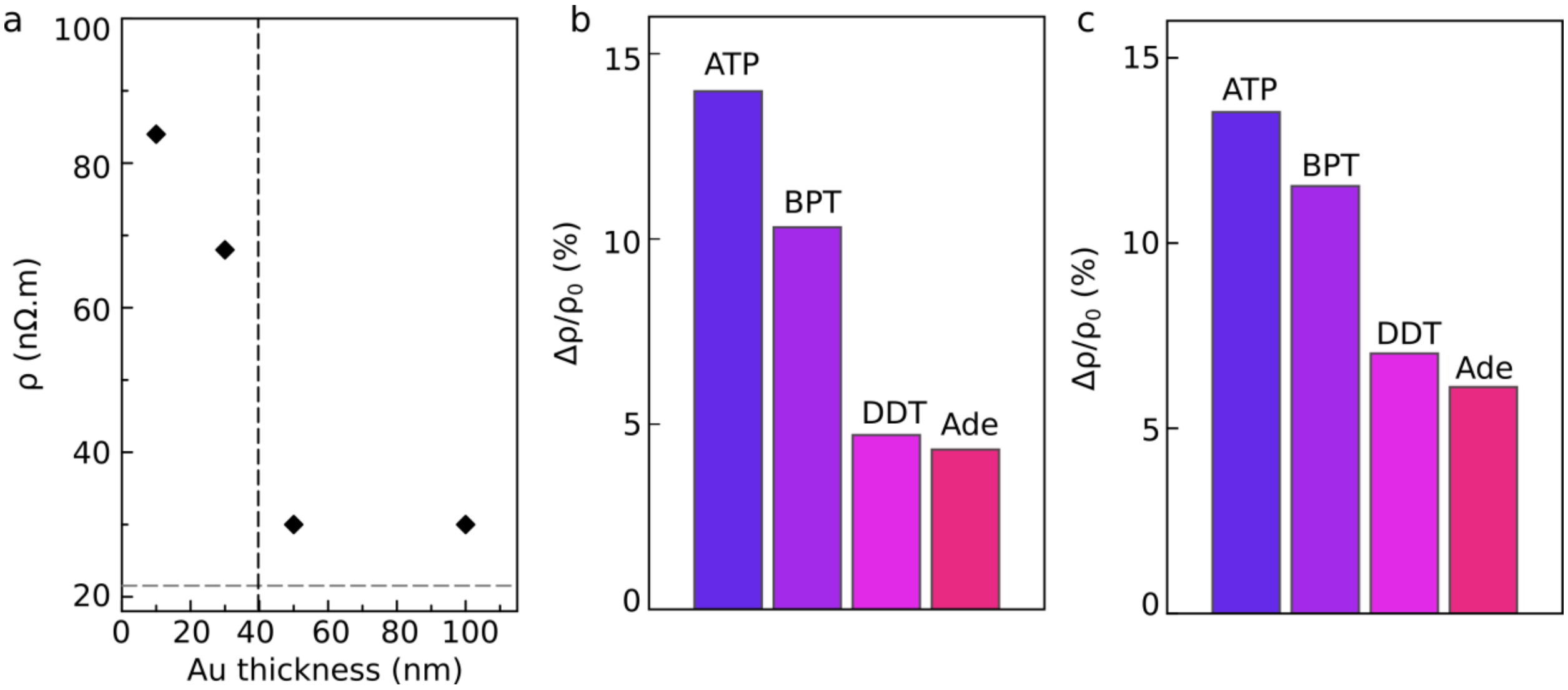

**Figure S2.** a) Resistivity of the blank Au films with different thickness. Relative change of resistivity of b) 30 and c) 15 nm thick Au films with the four different adsorbates upon overnight functionalization (1 mM concentration solutions).

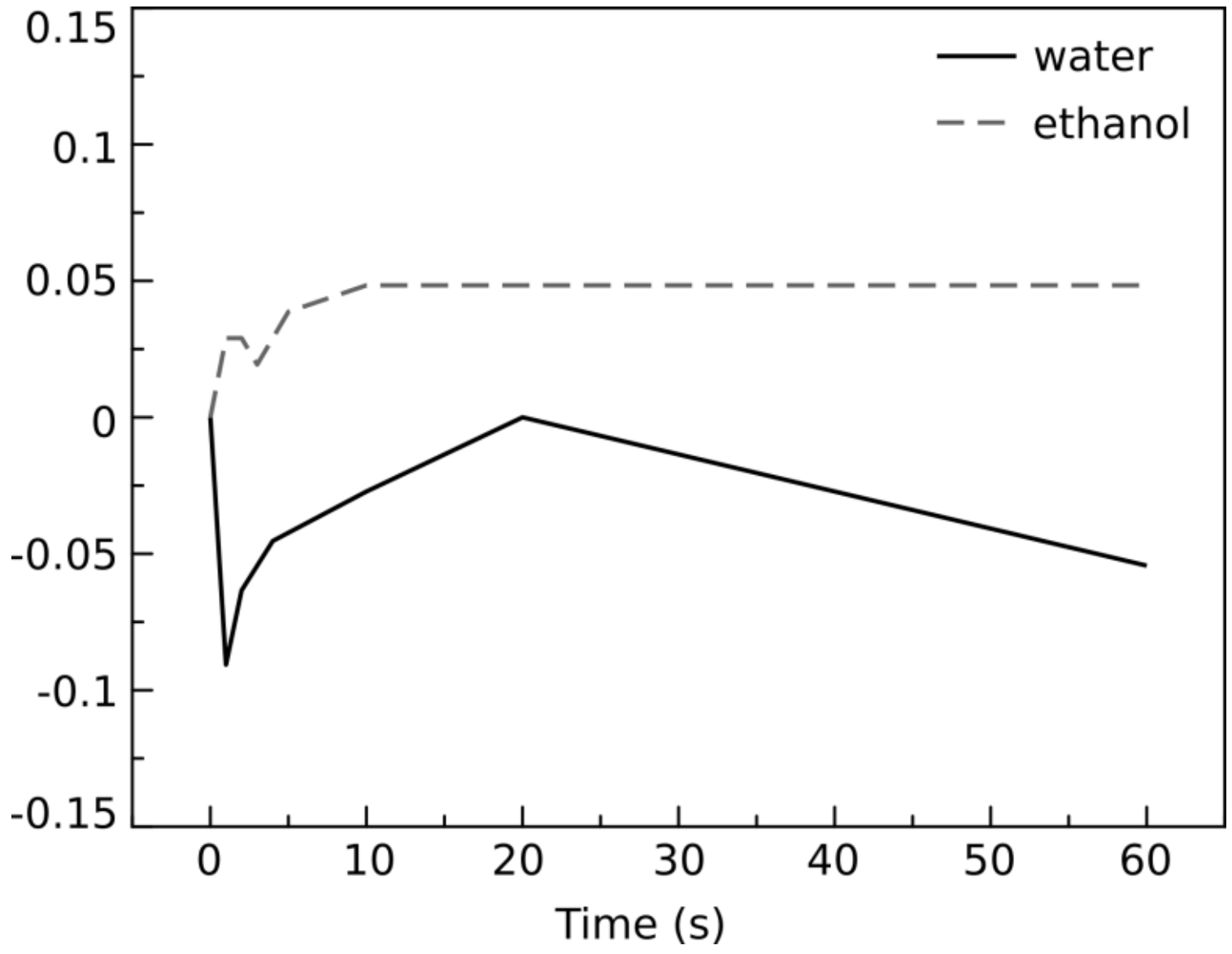

**Figure S3.** The relative change of Au film (30 nm) resistivity in time due to water and ethanol solvents only. Both solvents show a negligible change of resistivity. The slight decrease of resistivity for water is due to the solvation of physisorbed O atoms from the Au surface.

**Supplementary CID data**

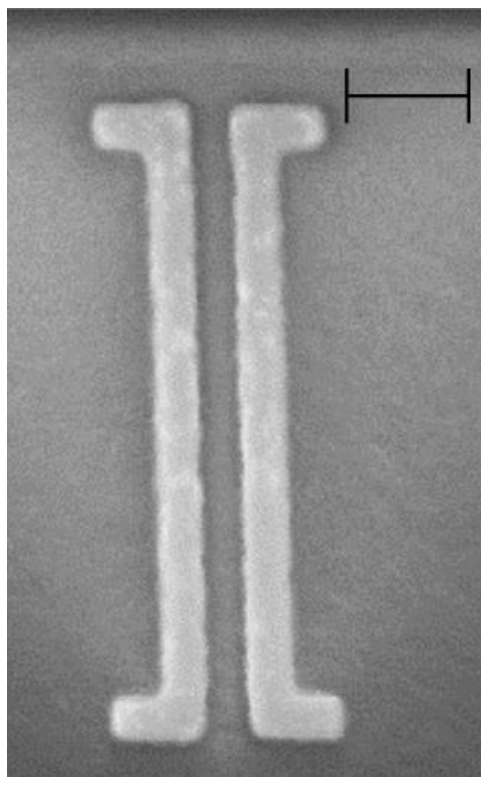

**Figure S4.** SEM image of the waveguide. The scale bar is 200 nm.

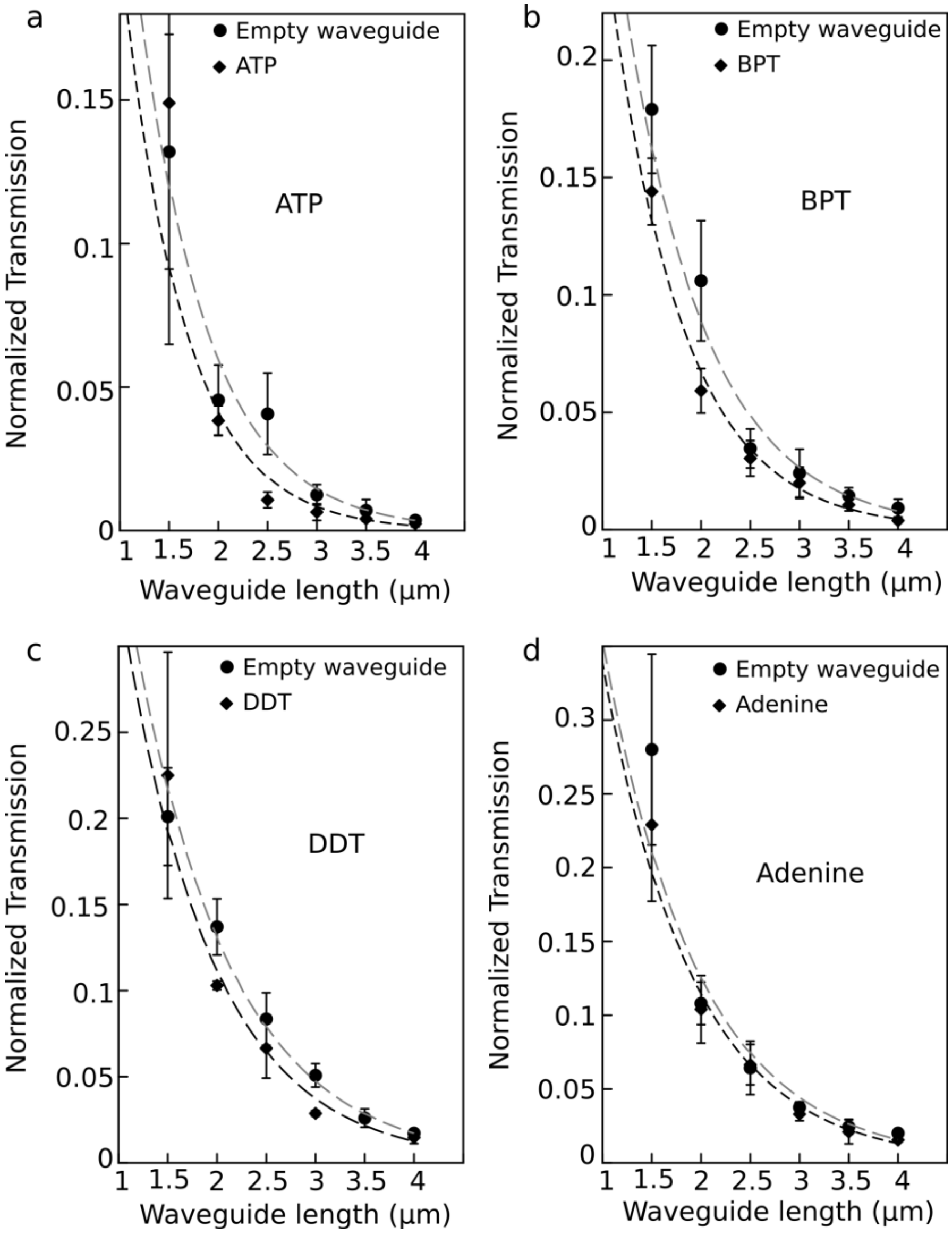

**Figure S5.** The total transmission of the waveguide versus the waveguide length with different molecular monolayers plotted. The data is normalized based on the assumption that the antenna efficiency is unchanged after molecule deposition.

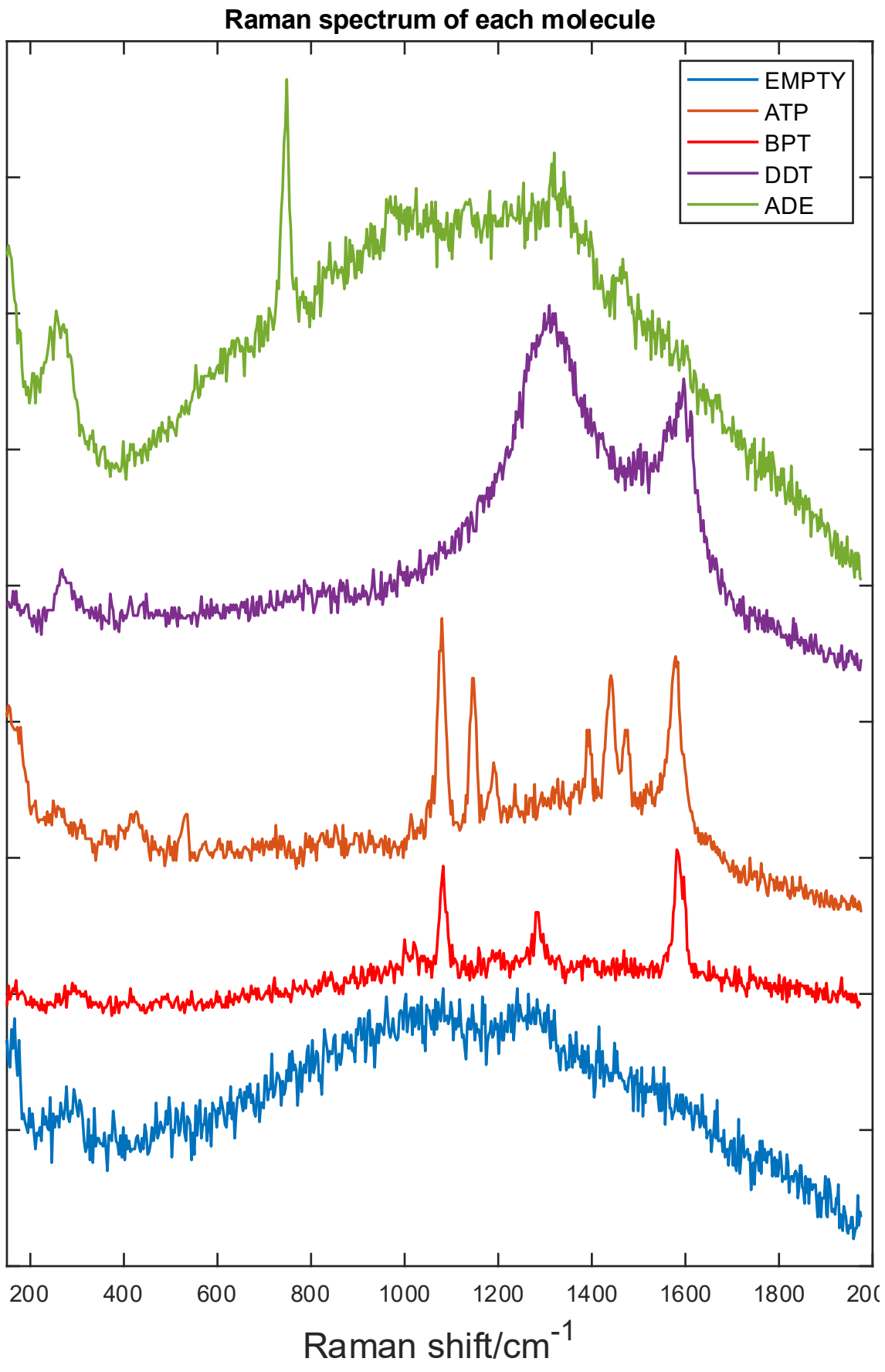

**Figure S6.** Raman spectrum of the empty plasmonic waveguide (blue spectrum) and the waveguide after molecular functionalization.

**Table S2.** Raw data for plasmon propagation loss on unfunctionalized waveguides and the waveguides with the four different adsorbates.

| Molecules | $L_{sp}$[µm] | $\alpha$ [µm$^{-1}$] | $\eta$ | $\alpha_{cid} = \Delta\alpha$ [µm$^{-1}$] | $\Delta\alpha/\alpha$ | $\gamma_{cid}$ [s$^{-1}$] |
|---|---|---|---|---|---|---|
| None | $0.83 \pm 0.19$<br>$0.71 \pm 0.05$<br>$0.97 \pm 0.03$<br>$0.96 \pm 0.07$* | $1.20 \pm 0.28$<br>$1.41 \pm 0.10$<br>$1.03 \pm 0.03$<br>$1.04 \pm 0.08$* | $0.20 \pm 0.01$<br>$0.16 \pm 0.03$<br>$0.31 \pm 0.03$<br>$0.14 \pm 0.04$ | – | – | – |
| BPT | $0.70 \pm 0.09$ | $1.43 \pm 0.04$ | $0.20 \pm 0.01$ | $0.23 \pm 0.33$ | $0.19 \pm 0.28$ | $2.6 \times 10^{13}$ |
| ATP | $0.63 \pm 0.02$ | $1.59 \pm 0.05$ | $0.16 \pm 0.06$ | $0.18 \pm 0.11$ | $0.12 \pm 0.08$ | $2.1 \times 10^{13}$ |
| DDT | $0.91 \pm 0.02$ | $1.10 \pm 0.02$ | $0.18 \pm 0.04$ | $0.07 \pm 0.05$ | $0.062 \pm 0.003$ | $8.1 \times 10^{12}$ |
| ADE | $0.91 \pm 0.06$ | $1.10 \pm 0.07$ | $0.20 \pm 0.03$ | $0.06 \pm 0.10$ | $0.05 \pm 0.01$ | $7.0 \times 10^{12}$ |

**Dipole moment of adsorbed molecules**

**Table S3. Calculated dipole moment of the four molecules in gas phase.**

| **Gas Phase** | **$\mu_x$ (D)** | **$\mu_y$ (D)** | **$\mu_z$ (D)** | **$\mu_{Total}$ (D)** |
|---|---|---|---|---|
| ATP | -6.49894 | 0.96882 | -1.52291 | 6.74493 |
| BPT | -0.29990 | 0.25831 | -4.61378 | 4.63073 |
| ADE | -0.12221 | 2.35480 | 0.33608 | 2.38180 |
| DDT | 1.33521 | -0.53598 | -1.41358 | 2.01699 |

**Table S4. Calculated dipole moment of the four molecules adsorbed on Au.**

| **Adsorbed** | **$\mu_x$ (D)** | **$\mu_y$ (D)** | **$\mu_z$ (D)** | **$\mu_{Total}$ (D)** |
|---|---|---|---|---|
| ATP | -1.97026 | 0.29265 | -3.15089 | 3.72769 |
| BPT | 0.74087 | 0.17044 | -5.68109 | 5.73173 |
| ADE | 0.25489 | 0.42529 | -1.20447 | 1.30253 |
| DDT | 0.16619 | -0.11374 | -2.85415 | 2.86125 |

## Computational Details

The four adsorbates we study are adenine (ADE), 4-aminothiophenol (ATP), biphenylthiol (BPT) and 1-dodecanethiol (DDT). We use a slab approach to model an Au (111) surface. The slab has 5 layers, a vacuum size of 10 *Å* and we construct a supercell of the slab.
For ADE and ATP we use a 3x3 supercell, and for BPT and DDT we use a $2\sqrt{3}\text{x}\sqrt{3}$ supercell, corresponding to different adsorbate densities. The geometry optimization was performed using the Vienna Ab Initio Simulation Package (VASP) with projector augmented waves (PAW) pseudopotentials and the PBE functional.[7-11] The plane wave cutoffs were chosen to be 420 *eV* for each structure. A Gaussian smearing scheme was used with a Gaussian broadening of 0.1 *eV*. Van der Waals corrections and dipole corrections were included. The Van der Waals correction was applied via the Grimme D3 dispersion correction.[12-13] All geometry optimizations were carried out at the Γ point, and the structures were relaxed to a force tolerance of 1 *eV/Å* per atom. The relaxed structures are shown in Fig. S7.

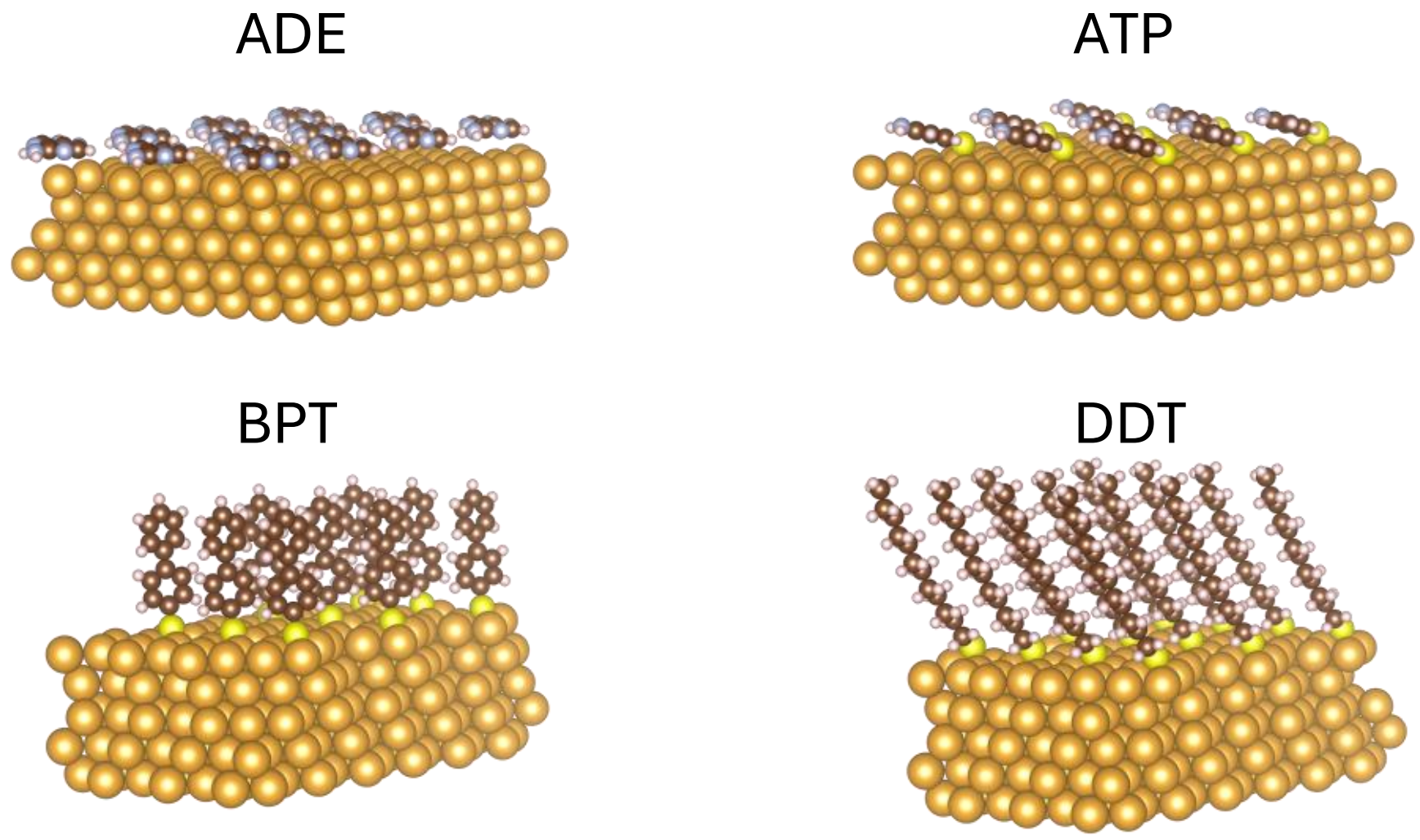

**Figure S7:** Relaxed structures of ADE, ATP, BPT and DDT adsorbed onto an Au (111) surface.
To evaluate the integrals from Persson's model, a density of states (DOS) must be obtained and projected onto the adsorbate.

Quantum Espresso was used for the density of states calculations.[14-16] GBRV ultrasoft pseudopotentials were used with a wavefunction cutoff of 40 *Ry* with a PBE functional.[17] A Gaussian smearing scheme was used with a broadening of 0.004 *Ry*. Both Grimme D3 Van der Waals dispersion correction and a dipole correction were applied. For the self-consistent field (SCF) cycles, a k-grid of 4x4x1 was used for each structure, and for the non self-consistent field cycles, a k-grid of 14x14x1 was used.
Persson's model uses a free electron gas model for Au. Therefore, a consistent choice for the Fermi energy that appears in Persson's integrals must be chosen. We choose to shift the DFT energies such that the Fermi energy for each structure agrees with the result from the free electron gas, i.e. $\varepsilon_F = \hbar^2(3\pi\text{n})^{\frac{2}{3}}/\text{m}_e$, where $\varepsilon_F$ is the Fermi energy of the free electron gas, $n$ is the number density of electrons and $m_e$ is the mass of the electron.

To calculate the induced dipole moment in the metal due to the presence of the adsorbate, we use a cluster model of the Au (111) system.

To create the cluster, we start with the optimized periodic slab structures, taking one supercell and adding additional gold atoms around it. A cutoff radius is then chosen from a point centered on a gold atom directly below the adsorbate. All atoms that fall within this cutoff radius are included in the cluster, and all others are discarded. The remaining cluster is then centered in a 40x40x40 $A^{\circ}$ box. This procedure is shown diagrammatically in Fig. S8. The dipole moment is long ranged and so to capture the effects of the induced dipole moment a large surface area is needed. Thus, to keep computational costs down, only 3 layers of Au atoms are used in the cluster calculations.

To model the induced dipole moment due to the adsorbate, we take away the dipole moment of a pristine Au (111) cluster and the dipole moment of an isolated adsorbate away from the dipole moment of the Au (111) cluster with the adsorbate.

DFT calculations were performed on the clusters from Fig. S7 using the Vienna Ab Initio Simulation Package (VASP) with projector augmented waves (PAW) pseudopotentials and the PBE functional. The wavefunction cutoff was chosen to be 420 *eV* for each calculation. For calculations containing Au, a Gaussian smearing scheme was used with a Gaussian broadening of 0.1 *eV*. Both Grimme D3 Van der Waals dispersion correction and a dipole correction were applied to all calculations. All calculations were performed at the Γ point. The dipole moment was then obtained from the charge density.

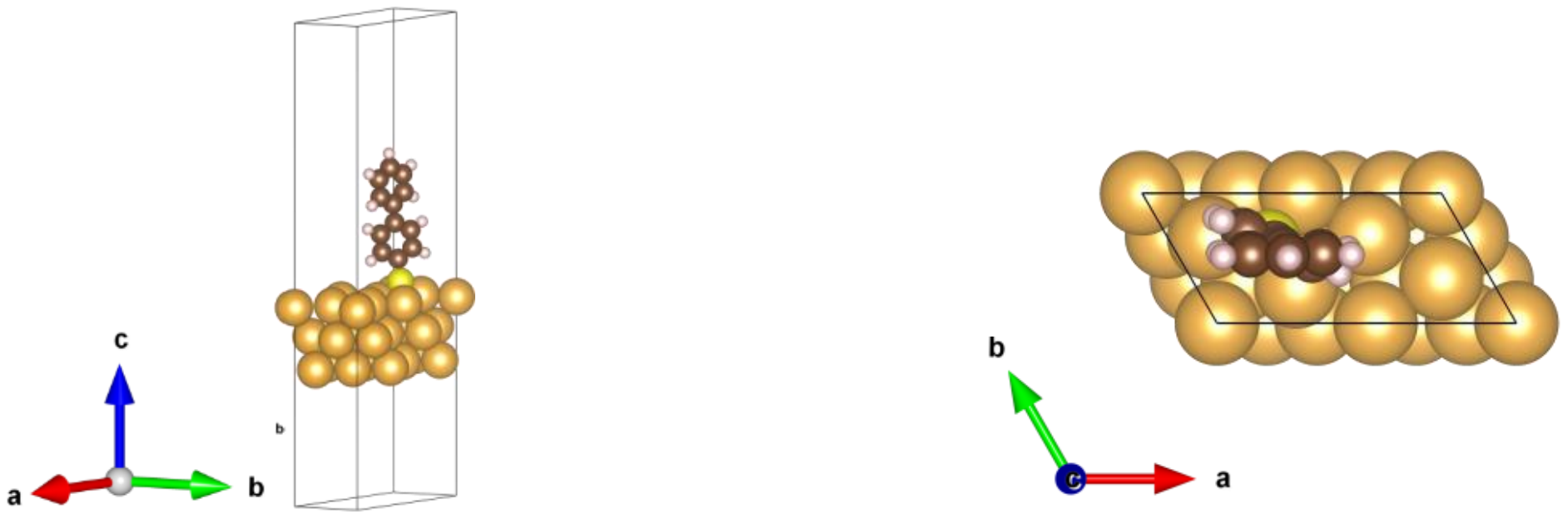

(a) Top view of optimized BPT periodic slab (b) Side view of optimized BPT periodic slab

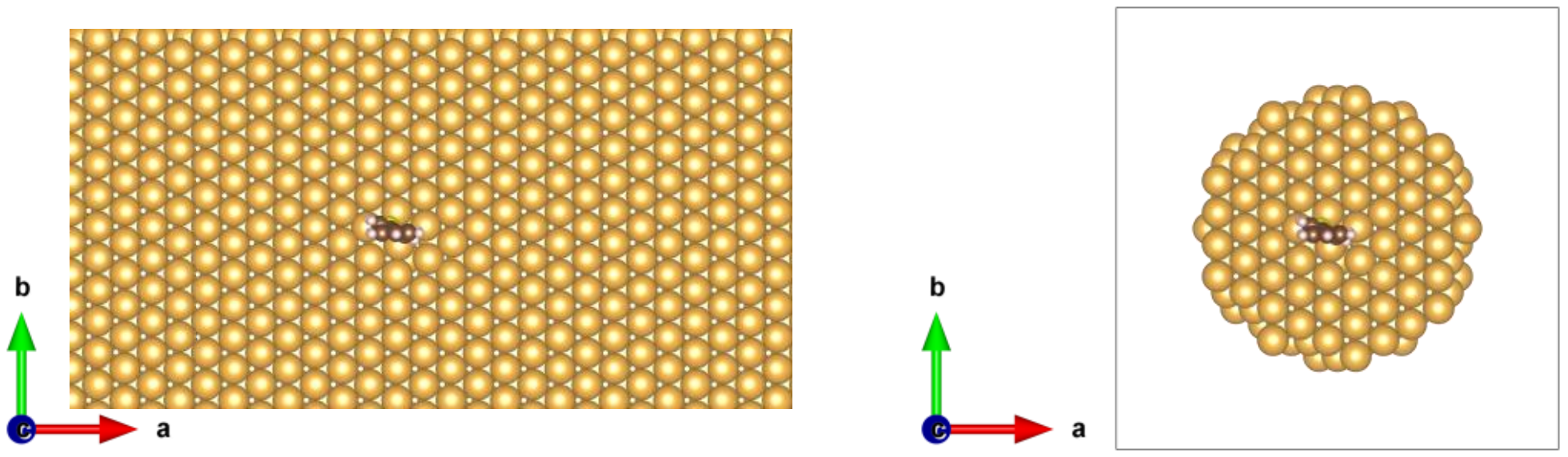

(c) Top view of optimized BPT periodic slab (d) Top view of cluster with extra Au atoms

**Figure S8**: A relaxed periodic unit cell with one adsorbate (a, b) has Au atoms added to the boarders of the unit cell to produce (c). A defined cutoff radius is then specified to carve a cluster out of the system, e.g. a 12 $\AA$ radius was chosen to make (d).